\documentclass[final]{elsart}
\usepackage{ifpdf,graphicx,epsfig,bm,amsfonts,amssymb,amsmath}
\newcommand{\GAMMA}{\mbox{\boldmath${\gamma}$}}

\newcommand{\be}{\begin{equation}}
\newcommand{\ee}{\end{equation}}
\newcommand{\bq}{\begin{eqnarray}}
\newcommand{\eq}{\end{eqnarray}}

\newcommand{\no}{\nonumber\\}

\newcommand{\1}{1\!\!1}

\begin{document}

\begin{frontmatter}

\title{Yang-Mills gauge theories from simple fermionic lattice models}

\author[it]{Paolo~Maraner} and
\author[uk]{Jiannis~K.~Pachos}

\address[it]{School of Economics and Management, Free University of Bozen-Bolzano,\\
             via Sernesi~1, 39100~Bolzano, Italy}
\address[uk]{School of Physics and Astronomy, University of Leeds, Leeds LS2 9JT, UK}

\begin{abstract}
A doublet of three-dimensional Dirac fermions can effectively describe the
low energy spectrum of a fermionic cubic lattice. We employ this fermion
doubling to encode a non-Abelian $SU(2)$ charge in the fundamental
representation. We explicitly demonstrate that suitable distortion of the
tunnelling couplings can introduce a scalar and a Yang-Mills field in the
effective low energy description, both coupled to the Dirac fermions. The
simplicity of the model suggests its physical implementation with ultra-cold
atoms or molecules.
\end{abstract}


\end{frontmatter}

\section{Introduction}
\label{Intro}

Recently, much interest has been focused on fermionic lattice models that
can simulate one or two-dimensional exotic phenomena such as
fractionalization of charge or anyonic statistics. Polyacetylene is a
one-dimensional example, which gave a simple and rigorous theoretical model
for the charge fractionalization~\cite{Jackiw0} that enjoyed experimental
verification~\cite{Kivelson}. In two-dimensions graphene, a mono-atomic
layer of graphite, has a low energy description given by a two-dimensional
Dirac equation~\cite{Wallace}. It has been shown that under certain coupling
distortions of the underlining lattice it can support axial gauge theories
that lead to the fractionalization of
charge~\cite{Chamon,Jackiw3,Marcel,Pachos}. In three or four dimensions the
properties of these lattices and their coupling to external non-Abelian
gauge fields are well understood, for example, in the context of staggered
fermion lattice gauge theories~\cite{Golterman,Stern,Mitra}. How non-Abelian
gauge fields can effectively arise in the continuum limit without employing
them already in the lattice picture remained up to now an open
question~\cite{Jackiw3,Volovik,Wen}.

In this letter, we systematically study the case of three spatial
dimensions. We employ a cubic lattice model where fermions are tunnelling
from site to site with specific transition couplings. It has been
shown~\cite{Susskind} that the low energy spectrum of this model is
effectively described by three-dimensional Dirac spinors. Here, we
demonstrate that suitable {\em distortions} of the transition couplings and
of the local potentials give rise to additional triplet scalar and
Yang-Mills fields, both of them coupled to the Dirac field. Importantly, the
presented distortions do not originate from non-Abelian fields in the
lattice picture. The interest in such structures is diverse. The simplicity
of these lattice models could inspire physical systems such as optical
lattices or multi-layer graphene that can simulate Dirac fermions coupled to
dynamical Yang-Mills gauge fields. Such an endeavor could establish a bridge
between the lattice gauge theory and the experimental condensed matter
communities. Moreover, it is of interest to see if the standard model can
emerge as an effective theory of a simple underlined lattice
structure~\cite{Wen}. Such a model could generate Abelian and non-Abelian
gauge potentials as well as Higgs scalar fields that emerge without a
symmetry breaking due to a simple quantum ordered in its ground state.

\section{The three dimensional case}

\subsection{The fermionic lattice model and the Dirac equation}

For concreteness we consider the three-dimensional Hamiltonian
\begin{eqnarray}
D\Psi=(\1 \otimes{\GAMMA}\cdot{\mathbf{p}} -T^a\otimes
{\GAMMA}\cdot{\mathbf{A}}_a+T^a\otimes\1\Phi_a )\Psi,
\label{Ham1}
\end{eqnarray}
where summations are assumed over repeated indices. The gauge field is taken
to be in the fundamental representation of $SU(2)$, i.e. $T^a={\sigma_a\over
2}$~\cite{comment}. The four spinor, $\Psi_i^{(n)}$, is indexed by
$i=1,...,4$, while $n=1,2$ runs trough the color components. The index
$a=1,2,3$ runs through the gauge field components and
$\{\gamma_\mu,\gamma_\nu\}= 2\delta_{\mu\nu}$, where $\mu$, $\nu$ run
through the space components $x$, $y$ and $z$. By choice of gauge the
three-dimensional gauge vector $\mathbf{A}_a$ and the triplet scalar field
$\Phi_a$ are taken to be real.

\begin{figure}
\begin{center}
\includegraphics[width=3.3cm,height=3.6cm]{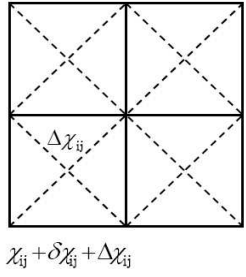}
\caption{A two-dimensional section of the three-dimensional cubic lattice.
Fermions, $\psi_\mathbf{i}$, tunnel from site $\mathbf{i}$ to site
$\mathbf{j}$ along the links of the lattice with couplings
$X_\mathbf{ij}=\chi_\mathbf{ij} +\delta\chi_\mathbf{ij}
+\Delta\chi_\mathbf{ij}$. Here $\chi_\mathbf{ij}$ and
$\delta\chi_\mathbf{ij}$ couple nearest neighboring sites, $\mathbf{i}$ and
$\mathbf{j}$, that lie only along the same link of the cubic lattice, while
$\Delta\chi_\mathbf{ij}$ couples diagonal sites as well.}
\label{fig:lattice}
\end{center}
\end{figure}

Our starting point is a cubic lattice where fermions tunnel from one site to
the neighboring one according to the simple quadratic Hamiltonian
\be
\hat H= -\sum_{\langle \mathbf{i,j}\rangle}
\psi_\mathbf{i}^\dagger X_\mathbf{ij} \psi_\mathbf{j} +{\rm H.c.}.
\label{Ham2}
\ee
The tunnelling couplings are given by
\be
X_\mathbf{ij} = \chi_\mathbf{ij}+\delta\chi_\mathbf{ij}
+\Delta\chi_\mathbf{ij},
\ee
where $\chi_\mathbf{ij}$ correspond to a uniform pattern, while
$\delta\chi_\mathbf{ij}$ and $\Delta\chi_\mathbf{ij}$ correspond to coupling
distortions that could have a slow spatial variation. While
$\chi_\mathbf{ij}$ and $\delta\chi_\mathbf{ij}$ connect nearest neighbor
sites that lie only along the links of the cubic lattice, the couplings
$\Delta\chi_\mathbf{ij}$ can also generate site energy shifts or couple
sites along diagonals, as seen in Fig.~\ref{fig:lattice}. As we shall see
the couplings $\chi_\mathbf{ij}$ are chosen so that the low energy
description of $\hat H$ is given by a Dirac equation. There are three
independent components of $\delta\chi_\mathbf{ij}$, along the different
directions of the lattice, that can give rise to a triplet scalar field,
while nine independent components of $\Delta\chi_\mathbf{ij}$ can be
identified that can give rise to the $SU(2)$ non-Abelian gauge potential.

Let us first consider the effect of the uniform couplings $\chi_\mathbf{ij}$
that give rise to the Kogut-Susskind (or staggered)
fermions~\cite{Susskind}. Their particular values that actually minimize the
energy of Hamiltonian (\ref{Ham2}) are given by
\be
\chi_{\mathbf{i,i+\hat x}} = -{i \over 2},\,\chi_\mathbf{i,i+\hat y} =
-{i\over 2}(-1)^{i_x},\,\chi_\mathbf{i,i+\hat z} = -{i\over
2}(-1)^{i_x+i_y}.
\nonumber
\ee
This configuration corresponds to a $\pi$-flux going through each square
plaquette of the lattice. As the couplings, $\chi_\mathbf{ij}$, have
alternating signs along the $x$ and $y$ directions, while they are
homogeneous along the $z$ direction, the resulting Brillouin zone, $B$, is
given by $k_x,k_y\in[-\pi/2,\pi/2]$ and $k_z\in[-\pi,\pi]$. In  momentum
space the Hamiltonian kernel, $H$, is given by
\be
H=\1\otimes\sigma_3 \sin k_x + \sigma_3 \otimes \sigma_1 \sin k_y +
\sigma_1\otimes\sigma_1 \sin k_z
\label{Ham3}
\ee
in the basis $(\mathbf{k, k+Q_x,k+Q_y,k+Q_x+Q_y})$, where
$\mathbf{Q_x}=(\pi,0,0)$ and $\mathbf{Q_y}=(0,\pi,0)$. The corresponding
eigenvalues of the energy are given by
\be
E_{\pm} =\pm \sqrt{\sin^2 k_x + \sin^2 k_y +\sin^2 k_z}.
\nonumber
\ee
Within the Brillouin zone, $B$, there are two isolated Fermi points given by
$\mathbf{K}_+ =(0,0,0)$ and $\mathbf{K}_-=(0,0,\pi)$ that set the energy,
$E_\pm$, to zero. Focusing on the low energy regime of the Hamiltonian at
half filling we expand (\ref{Ham3}) around $\mathbf{K}_+$ and
$\mathbf{K}_-$. For $\mathbf{k}=
\mathbf{K}_\pm +
\mathbf{p}$ we have
\be
H_\pm=\1\otimes \sigma_3 p_x + \sigma_3 \otimes \sigma_1 p_y
\pm\sigma_1\otimes\sigma_1 p_z,
\label{Ham4}
\ee
where $|\mathbf{p}|$ is taken to be small. We can now define
$\gamma_x=\1\otimes
\sigma_3$, $\gamma_y=\sigma_3
\otimes
\sigma_1$ and $\gamma_z=\sigma_1\otimes\sigma_1$ so that $H_+=
\GAMMA\cdot \mathbf{p}$. The corresponding spinors are given by
$\Psi_\pm(\mathbf{p})^T=(\psi_{\pm,\mathbf{p}},\psi_{\pm,\mathbf{p}+
\mathbf{Q}_x},\psi_{\pm,\mathbf{p}+\mathbf{Q}_y}, \psi_{\pm,\mathbf{p}+
\mathbf{Q}_x+\mathbf{Q}_y})$ where $\psi_{\pm,\mathbf{p}}=
\psi(\mathbf{K}_\pm+\mathbf{p})$. Since $(\sigma_3\otimes\1)H_-(\sigma_3\otimes
\1)=H_+$ it is easy to show that for $U=\text{diag}(\1\otimes\1
,\sigma_3\otimes\1)$ it is
\be
U\left(
\begin{array}{cc}
H_+ & 0 \\
0 & H_- \\
\end{array}
\right)U^\dagger=\1\otimes{\GAMMA}\cdot \mathbf{p}.
\ee
In this way one can obtain two identical copies of the free
three-dimensional Dirac equation. The spinor, $\Psi$, of eqn.~(\ref{Ham1})
is given by $\Psi^{(1)}_i=U{\Psi_+}_i$ and $\Psi^{(2)}_i=U{\Psi_-}_i$, where
$i=1,...,4$ enumerates the four components of the spinors.

\subsection{The scalar field}

In order to obtain the scalar field of Hamiltonian~(\ref{Ham1}) we choose
the coupling distortions to be of the form
\be
\delta \chi_{\mathbf{i,i+\hat y}}=-{\Phi \over 4} e^{-i\mathbf{K}_-\cdot\mathbf{i}}
e^{i\mathbf{K}_+\cdot(\mathbf{i}+\mathbf{s}_y)},\,\,\delta
\chi_\mathbf{i,i+\hat z}=-{\Phi_3 \over 8},
\ee
where $\Phi=\Phi_1-i\Phi_2$ with all $\Phi_a$ real. The $\mathbf{i}$ vector
indicates the position of the corresponding site and the $\mathbf{s}_a$
vector indicates its two nearest neighbors along the direction $a$. The
components of the scalar field can have a variation in space, which is slow
compared to the lattice spacing. As $\delta\chi_\mathbf{i,i+\hat x} =0$ we
have that $\delta H^{(x)}=0$. For the distortion
$\delta\chi_\mathbf{i,i+\hat y}$ and for momenta close to the Fermi points
we obtain the additional Hamiltonian term
\be
\delta H^{(y)} = {\Phi_1 \sigma_1 +\Phi_2 \sigma_2\over 2}\otimes \sigma_3
\otimes \1.
\nonumber
\ee
The distortion in the $z$ direction, $\delta \chi_\mathbf{i,i+\hat z}$,
gives the Hamiltonian contribution
\be
\delta H^{(z)} = {\Phi_3 \over 2} \sigma_3\otimes\1\otimes\1.
\nonumber
\ee
Finally, the conjugation $U(\delta H^{(y)}+\delta H^{(z)})U^\dagger$ results
to the triplet scalar field term, $T^a\otimes\1\Phi_a$, of
Hamiltonian~(\ref{Ham1}).

\subsection{The Yang-Mills field}

The gauge field can originate from the $\Delta\chi_\mathbf{ij}$ distortion.
The non-zero couplings are given by
\bq
&&\Delta\chi_\mathbf{i,i+\hat y}=(-1)^{i_x+i_y} {A^z
\over 4} e^{-\mathbf{K}_-\cdot\mathbf{i}} e^{i\mathbf{K}_+\cdot\mathbf{i}},
\nonumber \\
&&\Delta\chi_\mathbf{i,i}=-(-1)^{i_x+i_y}{A^z_3 \over 4}-(-1)^{i_x}{A^y
\over 4} e^{-i\mathbf{K}_-\cdot\mathbf{i}} e^{i\mathbf{K}_+\cdot\mathbf{i}},
\no
&&\Delta\chi_\mathbf{i,i+\hat x+\hat y}=-{A^x \over 8}
e^{-i\mathbf{K}_-\cdot\mathbf{i}} e^{i\mathbf{K}_+\cdot\mathbf{i}},
\no
&&\Delta\chi_\mathbf{i,i+\hat x+\hat z} =-{A_3^x \over 16},\,\,
\Delta\chi_\mathbf{i,i+\hat y+\hat z} = -(-1)^{i_x}{A_3^y \over 16},
\eq
where $A^\mu=A_1^\mu-iA_2^\mu$ with all $A_a^\mu$ real and slowly varying
with respect to the lattice spacing. These distortions of the tunnelling
couplings provide the nine components of the non-Abelian gauge potential.
Indeed, $\Delta\chi_\mathbf{i,i+\hat y}$ gives rise to
\be
\Delta H^{(y)} = -{A_1^z\sigma_2-A_2^z\sigma_1\over 2}
\otimes\sigma_2\otimes\sigma_1.
\nonumber
\ee
From $\Delta\chi_\mathbf{i,i}$, which is a distortion of the local potential
of site $\mathbf{i}$, we obtain
\be
\Delta H^\mathbf{(0)} = -{A_3^z \over 2} \1\otimes\sigma_1\otimes\sigma_1 -
{A_1^y\sigma_1 + A_2^y\sigma_2\over 2}\otimes\1\otimes\sigma_1.
\nonumber
\ee
From $\Delta\chi_\mathbf{i,i+\hat x+\hat y}$ we obtain the additional
Hamiltonian term
\be
\Delta H^\mathbf{(\hat x+\hat y)} = -{A_1^x\sigma_1+A^x_2\sigma_2\over 2}
\otimes \sigma_3\otimes\sigma_3,
\nonumber
\ee
while from $\Delta\chi_\mathbf{i,i+\hat x+\hat z}$ we obtain
\be
\Delta H^\mathbf{(\hat x+\hat z)} = -{A_3^x \over 2}\sigma_3\otimes
\1\otimes\sigma_3.
\nonumber
\ee
Finally, from $\Delta\chi_\mathbf{i,i+\hat y+\hat z}$ we obtain
\be
\Delta H^\mathbf{(\hat y+\hat z)} = -{A_3^y\over 2} \sigma_3\otimes
\sigma_3\otimes\sigma_1.
\nonumber
\ee
After conjugating by $U$ we have that $U(\Delta H^{(y)}+\Delta
H^\mathbf{(0)}+\Delta H^\mathbf{(\hat x+\hat y)}+\Delta H^\mathbf{(\hat
x+\hat z)} +\Delta H^\mathbf{(\hat y+\hat z)})U^\dagger
=-T^a\otimes\GAMMA\cdot\mathbf{A}_a$. This shows that the
$\Delta\chi_\mathbf{ij}$ coupling distortions followed by the $U$
conjugation give rise to the gauge potential term of
Hamiltonian~(\ref{Ham1}). Thus, by appropriate distortions of the tunnelling
couplings of the fermionic lattice model one can generate a non-Abelian
gauge and a scalar field coupled to a Dirac field. Note, that, while these
distortions may have a complex value the total {\em additional} plaquette
flux they introduce is zero. Thus, the initial lattice
Hamiltonian~(\ref{Ham2}) is time symmetric.

\subsection{Yang-Mills monopoles and the index theorem}

In our construction the gauge and scalar field components can take arbitrary
configurations. As an application, we would like to simulate Yang-Mills
monopoles in the low energy behavior of the presented fermionic lattice
model by appropriately choosing the coupling distortions,
$\delta\chi_\mathbf{ij}$ and $\Delta \chi_\mathbf{ij}$. The field
configurations, induced by a single monopole positioned at the origin with
unit charge, are given by
\be
\Phi_a = \hat r_a \Phi(r),\,\,A_a^\mu = \epsilon^{a\mu\nu}\hat r_\nu A(r).
\label{monopole}
\ee
Both $\Phi(r)$ and $A(r)$ are taken to vanish at $r=0$, while for large $r$
the scalar field tends to its vacuum value,
$\Phi\xrightarrow[r\rightarrow\infty]{}m$, and the gauge field behaves as
$A\xrightarrow[r\rightarrow\infty]{}-1/r$, where the approach to their
asymptotic behavior is exponential~\cite{tHooft}. Substituting these
configurations into Hamiltonian~(\ref{Ham1}) one obtains
\be
D\Psi=\big[\1 \otimes{\GAMMA}\cdot{\mathbf{p}} -{\sigma_a\over 2}
\otimes ({\GAMMA}\times
\hat{\mathbf{r}})_aA + {\sigma_a\over 2}\otimes\1\hat{r}_a\Phi\big] \Psi,
\label{Ham4}
\ee
which describes the Yang-Mills monopole coupled to a Fermi field.

The appearance of zero fermionic modes is the most striking characteristic
of such non-trivial topological configurations~\cite{Follana}. One can
employ the index theorem for open spaces to estimate the number of these
zero energy modes~\cite{Callias,Bott}. The index of an operator of the form
\be
D=\left(
\begin{array}{cc}
0 & L \\
L^\dagger & 0 \\
\end{array}
\right)
\label{HamInd}
\ee
is defined by $\text{index}(D)=k_+-k_-$, where $k_+$ ($k_-$) is the number
of zero modes of $L$ ($L^\dagger$). Its absolute value gives a lower bound
to the total number of zero modes, $k_++k_-$, of $D$. It is possible to
bring Hamiltonian~(\ref{Ham4}) in the form~(\ref{HamInd}) by conjugating it
with $\1\otimes[S(\1\otimes e^{i\sigma_1\pi/4})]$, where $S(\sigma_a\otimes
\sigma_b)=\sigma_b\otimes \sigma_a$ for all the Pauli matrices
including $\sigma_0=\1$. Then the index of Hamiltonian~(\ref{Ham4}) is given
by
\be
\text{index}(D)=-{1 \over 8\pi}
\int_{S^2_\infty}dS^\mu\epsilon^{\mu \alpha \beta}\epsilon^{abc}
{\Phi_c\partial_\alpha\Phi_a\partial_\beta\Phi_b \over |\Phi|^3},
\ee
which corresponds to the degree of $\Phi_a$ if it is considered as a mapping
from $S_\infty^2$ to $S_m^2$. For the single Yang-Mills monopole given
in~(\ref{monopole}) with charge $1$ we have $\text{index}(D) = 1$ indicating
that the system has (at least) one zero mode. Similarly, it is possible to
simulate higher charge monopoles, the explicit forms of which are given
in~\cite{Ward}.

It has been demonstrated~\cite{Jackiw} by analytic calculations that the
monopole configuration~(\ref{monopole}) possesses a unique normalized,
isolated and non-degenerate zero-energy mode. Its wave function is given by
$\Psi^{(n)}_{E=0}=(\phi^{(n)},0)^T$, where $\phi^{(n)}$ is a two component
spinor
\bq
\phi^{(n)}_{i}(r)=&&Ni\exp(-\int_0^r dr'\big[{1\over 2}
\Phi(r')+A(r')\big])
(\delta_{1i}\delta_{2n}-\delta_{2i}\delta_{1n})
\label{spinor}
\eq
with $i=1,2$ and $N$ a real normalization constant. This solution
corresponds to a wave function that is well localized around the position of
the monopole, $r=0$. The presence of a single zero mode in the spectrum of
the Dirac operator, together with the symmetry under fermi-number
conjugation, $\Psi^{(n)c}\equiv\sigma_3\otimes\sigma_2
(\sigma_2)_{nm}\Psi^{(m)*}$, of Hamiltonian~(\ref{Ham4}) signals the
fractionalization of the fermion number~\cite{Jackiw1}. Thus, the presence
of a classical topological defect in the configurations of the Yang-Mills
and scalar fields causes the fermionic field to take fractional quantum
numbers. This can be detected in the lattice fermion system by comparing the
fermion densities with and without the monopole. A fermion density
difference that corresponds to the localized wave function~(\ref{spinor})
provides a clear signature of the charged fractionalization.

\section{The two dimensional case}

For completeness we would like to present the generation of the Yang-Mills
field in the two-dimensional case. For that consider a square lattice with
tunnelling fermions governed by a similar Hamiltonian as in~(\ref{Ham1}).
The transition couplings are given by $X_\mathbf{ij}=
\chi_\mathbf{ij}+\Delta\chi_\mathbf{ij}$, with
\be
\chi_\mathbf{i,i+\hat x} =-{i \over 2},\,\,\chi_\mathbf{i,i+\hat y}=-{i\over
2} (-1)^{i_x}
\nonumber
\ee
and
\bq
&&\Delta \chi_\mathbf{i,i} = (-1)^{i_x}{A^y_3 \over 4} + {A^x \over
2}e^{-i(\mathbf{K_+-K_-})\cdot \mathbf{i}},
\no
&&\Delta \chi_\mathbf{i,i+\hat x+\hat y} = {A^x_3 \over 4},
\no
&&\Delta \chi_\mathbf{i,i+\hat x} = (-1)^{i_x} {A^y\over 2}
e^{-i(\mathbf{K_+-K_-})\cdot\mathbf{i}},
\eq
where $A^\mu=A_1^\mu-iA_2^\mu$, the Fermi points are $\mathbf{K}_+=(0,0)$
and $\mathbf{K}_-=(0,\pi)$ and the Brillouin zone, $B$, is given by
$k_x\in[-\pi/2,\pi/2]$ and $k_y\in[-\pi,\pi]$. The homogeneous tunnelling
couplings $\chi_\mathbf{ij}$ give rise to a pair of two-dimensional Dirac
equation corresponding to the two Fermi points. Under the conjugation by $U
=\text{diag}(\1,\sigma_3)$ the composite Hamiltonian becomes
$\1\otimes\GAMMA\cdot\mathbf{p}$, where $\GAMMA=(\sigma_3,\sigma_1)$. The
distortion $\Delta \chi_\mathbf{i,i}$ of the local site energies gives rise
to
\be
\delta H^\mathbf{(0)} = -{A^y_3 \over 2} \1 \otimes\sigma_1
-{A^x_1\sigma_1+ A^x_2\sigma_2 \over 2}\sigma_1\otimes\1.
\nonumber
\ee
From the distortion $\Delta \chi_\mathbf{i,i+\hat x+\hat y}$ we obtain
\be
\Delta H^{(\mathbf{\hat x+\hat y})} = -{A^x_3 \over 2} \sigma_3\otimes
\sigma_3
\nonumber
\ee
while from $\Delta \chi_\mathbf{i,i+\hat x}$ we have
\be
\Delta H^{(\mathbf{\hat x})} = - {A^y_1 \sigma_2- A^y_2\sigma_1 \over 2}
\otimes\sigma_2.
\nonumber
\ee
It is easy now to show that under conjugation by $U$ the overall Hamiltonian
produces the two dimensional Dirac operator coupled to an $SU(2)$ Yang-Mills
field $D\Psi = (\1\otimes\GAMMA\cdot\mathbf{p}  -T^a \otimes
\GAMMA \cdot\mathbf{A}_a)\Psi$.

It is an exciting possibility that such a lattice Hamiltonian could be
generated with optical lattices and ultra-cold atom or molecule
technology~\cite{Zoller}. In particular, the generation of the $\pi$-flux
through the square plaquettes of optical lattices is possible by imprinting
phases with laser assisted tunnelling processes~\cite{Jaksch}. The presented
method for generating Yang-Mills fields has the advantage that it does not
employ the internal states of the atoms that demand additional laser
fields~\cite{Osterloh} and, thus, it requires a significantly simpler
experimental setting for its physical realization. For example, the
distortion of the tunnelling couplings is expected to appear naturally in
optical lattice systems due to amplitude variations of the corresponding
laser fields. Alternatively, controlled spatially profiles of laser fields
can be easily produced in the laboratory without increasing the complexity
of the experiment. Applications of non-Abelian fields that facilitate the
robust simulation of the quantum Hall effect with ultra-cold atoms have been
reported in~\cite{Goldman}.

\section{Conclusions}

In conclusion, we considered a fermionic system with discretized three
spatial dimensions described by the Lagrangian
\be
\mathcal{L} = \sum_\mathbf{i} \psi_\mathbf{i}^\dagger i(\partial_t+iA^0)
\psi_\mathbf{i} +(\sum_\mathbf{\langle ij\rangle}
\psi_\mathbf{i}^\dagger X_\mathbf{ij} \psi_\mathbf{j}+\text{H.c.}),
\nonumber
\ee
where time is considered to be a continuous variable. By a particular choice
of the  transition couplings of the fermions we demonstrated that the low
energy limit of this system is faithfully described by Dirac fermions
coupled to a triplet scalar field, also known as a Higgs field, and a
Yang-Mills $SU(2)$ non-Abelian gauge field. These fields can be chosen to
encode Yang-Mills-Higgs monopole configurations. In particular, a single
monopole with charge $1$ can give rise to an isolated zero-mode. We
demonstrated that such a setting gives rise to fractionalization of charge
in three dimensions. Moreover, we derived the analytic relation between the
lattice fermions wave function and the Dirac ones, thus making it possible
to detect the topological effect of the effective gauge theory by probing
the behavior of the constituent lattice particles.  It is intriguing to see
if such systems in two or three dimensions can be simulated in the
laboratory.

The presented simulation opens up numerous possibilities. It would be
interesting to generate kinetic terms for the Yang-Mills field and augment
it to an independent dynamical entity in a quantum field theory. In this
context the study of the full symmetry group of the lattice and the
emergence of the correct massless spectrum in the quantum continuum limit is
highly desirable. Alternatively, a proof of stability of the quantum order
of these non-Abelian systems as well as the encoding of chiral fermions
would bring this effort closer into a full blown simulation of the standard
model.

\section*{Acknowledgments}
We thank Roman Jackiw and So-Young Pi for inspiring conversations throughout
this project. We would like to acknowledge the hospitality of the Aspen
Center for Physics. This work was supported by the EU grants EMALI and
SCALA, EPSRC and the Royal Society.


\begin{thebibliography}{99}
\bibitem{comment} Here $\sigma_a$, $a=1,2,3$ are the
Pauli matrices and $\sigma_0=\1$ is the two dimensional identity matrix.
\end{thebibliography}
\end{document}